\documentclass[twoside]{ilcws08}
\usepackage[latin1]{inputenc}
\usepackage[dvips]{graphicx,epsfig,color}
\usepackage{wrapfig,rotating}
\usepackage{amssymb,amsmath,array}

\newcommand{\sx}{$\sigma_{\mathrm{X}}$}
\newcommand{\sy}{$\sigma_{\mathrm{Y}}$}
\newcommand{\sz}{$\sigma_{\mathrm{Z}}$}

\newcommand{\mm}{mm$^{2}$}
\newcommand{\umum}{$\mu$m$^{2}$}
\newcommand{\um}{$\mu$m}

\begin{document}
\title{Development of Pair Monitor} %% 
%***********************************************************************
% AUTHORS INFORMATION AREA
%***********************************************************************
\author{Yosuke Takubo$^1$, Hirokazu Ikeda$^2$, Kazutoshi Ito$^{1}$, Akiya Miyamoto$^{3}$, Tadashi Nagamine$^{1}$,\\ Rei Sasaki$^{1}$, Toshiaki Tauchi$^{3}$, Yutaro Sato$^{1}$ and Hiroshi Yamamoto$^{1}$
% Optional short acknowledgment: remove next line if non-needed
%\thanks{This is an optional funding source acknowledgment.}
% DO NOT MODIFY THE FOLLOWING '\vspace' ARGUMENT
\vspace{.3cm}\\
% Addresses and institutions (remove "1- " in case of a single institution)
1- Department of Physics, Tohoku University \\
Research Center for Neutrino Science, Tohoku University, Sendai, Japan
%% Remove the next three lines in case of a single institution
\vspace{.1cm}\\
2- Institute of Space and Astronautical Science, \\
Japan Aerospace Exploration Agency (JAXA), Sagamihara, Japan 
\vspace{.1cm}\\
3- High Energy Accelerator Research Organization (KEK), Tsukuba, Japan
}
%%***********************************************************************
% END OF AUTHORS INFORMATION AREA
%***********************************************************************

\maketitle

\begin{abstract}
The pair monitor is a beam profile monitor at interaction point (IP) 
for the international linear collider (ILC). 
We have designed and developed the pair monitor as a silicon pixel sensor 
which is located at about 400 cm from IP. 
As the first step to develop the pair monitor, the readout ASIC was developed.
In this paper, test results of the readout ASIC and the future plan are reported.
\end{abstract}

%%%%%%%%%%%%%%%%%%%%%%%%%%%%%%%%%%%%%%%%%%
\section{Introduction}
%%%%%%%%%%%%%%%%%%%%%%%%%%%%%%%%%%%%%%%%%%

\begin{wrapfigure}{r}{0.5\columnwidth}
\centerline{\includegraphics[width=0.45\columnwidth]{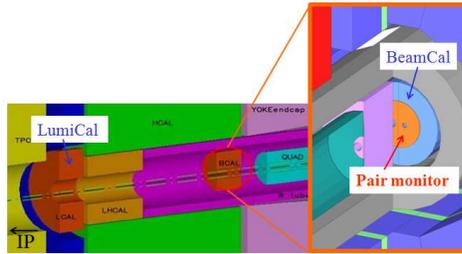}}
\caption{A schematic view of the forward region.
The pair monitor will be located in front of the BeamCal,
where is about 400 cm from IP.}
\label{fig:setup}
\end{wrapfigure}

At ILC, measurement of the beam profile at IP is important to 
keep high luminosity of $2 \times 10^{34}$ cm$^{-2}$s$^{-1}$.
The beam size at IP is 639 nm, 5.7 nm, and 300 $\mu$m
for horizontal (\sx), vertical (\sy), and longitudinal beam size (\sz),
respectively.
Since the beam size, especially \sy, is very small, 
the beam profile monitor is required to measure the beam size within 1 nm accuracy.

The pair monitor will be used to check the beam profile 
at the interaction point (IP),
measuring the distribution of the electron-positron pairs generated 
during the beam crossing \cite{tauchi}.
The generated electrons and positrons are scattered 
by the magnetic field produced by the oncoming beam, 
which is a function of the transverse size (\sx, \sy) 
and intensity of the beam. 
For that reason, 
the deflected particles should carry information of the transverse beam size, 
especially in their angular distribution. 
The pair monitor will be located at 400 cm from IP as shown in Fig. \ref{fig:setup}, 
where is in front of the BeamCal \cite{bcal}.
In our previous studies, 
the pair monitor has performance to measure the beam size with about 10\% accuracy \cite{takubo}.

We have studied design of the pair monitor 
and developed the prototype of the readout ASIC.  
In this paper, test results of the readout ASIC and the future plan are reported.

%The particles of concern have the same charge as that of the oncoming beam, 
%and are hereafter called "same-charge" particles. 
%Most of them are deflected at larger angles than their inherent scattering angles 
%by a strong electromagnetic force due to the oncoming beam, 
%while the "opposite-charge" particles must oscillate inside the oncoming beam 
%because of a focusing force between them; 
%they are deflected with small angles. 
%They can be well described by a scattering process of $e^{-}$($e^{+}$) 
%in a two-dimensional Coulomb potential 
%that is Lorents boosted to the rest frame of the oncoming beam. 
%Since this potential is produced 
%by the intense electric charge of the oncoming beam, 
%it is a function of the transverse size (\sx, \sy) 
%and intensity of the beam. 
%For that reason, 
%the deflected particles should carry information of the transverse beam size, 
%especially in their angular distribution. 
%The pair monitor uses this principle 
%to extract information of the beam profile at IP. 

%%%%%%%%%%%%%%%%%%%%%%%%%%%%%%%%%%%%%%%%%%%%%
\subsection{Design concept of pair monitor}
%%%%%%%%%%%%%%%%%%%%%%%%%%%%%%%%%%%%%%%%%%%%%
There are some requirements to the pair monitor to be used 
for the beam profile monitor at ILC. 
The pair monitor is required to measure the beam profile at the interaction point, 
and the measurement results must be feedback to the next train 
to keep the high luminosity. 
For that reason, the pair monitor must measure the hit distributions for train by train, 
and the data should be readout within the inter-train time ($\sim200$ ms). 
Since the pair monitor will be put at about 400 cm from IP 
and close to the beam pipe, the radiation dose on the pair monitor becomes large. 
For example, the radiation dose was estimated as about 10 Mrad/year 
at the radius of 1.8 cm from the extraction beam pipe in the GLD geometry \cite{glddod}. 
Although the radiation dose decreases rapidly for the larger radius, 
the pair monitor should have radiation tolerance above 1 Mrad/year. 

\begin{wrapfigure}{r}{0.5\columnwidth}
\centerline{\includegraphics[width=0.45\columnwidth]{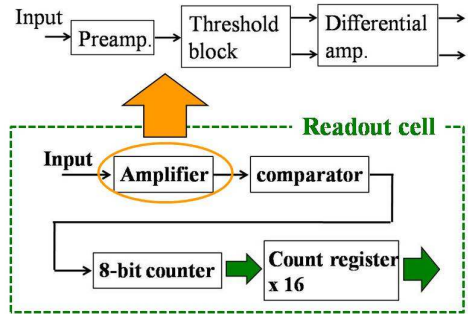}}
\caption{A schematic logic in the readout cell which consists of the amplifier,
comparator, 8-bit counter, and 16 count registers.
In the amplifier blocks, there are a pre-amplifier, threshold block, 
and differential amplifier.}
\label{fig:logic}
\end{wrapfigure}

To achieve these requirements, 
the design concept of the pair monitor was considered. 
The sensor is assumed as a silicon pixel sensor whose pixel size is 
$400 \times 400$ \umum~and thickness is about 200 \um. 
The size of the sensor layer is 10 cm radius. 
In the sensor layer, two holes for the injection 
and extraction beam pipes will be prepared, 
whose radius is 1.0 cm and 1.8 cm, respectively. 
The total readout channel will be about 200,000. 
The readout ASIC will be bump-bonded to the sensor, 
and measures the hit counts on the detector 
to obtain the hit distributions of the pair backgrounds. 
At that time, it is not necessary to obtain the information of the energy deposit. 
Based on this design concept, development of the pair monitor was started.

%%%%%%%%%%%%%%%%%%%%%%%%%%%%%%%%%%%%%%%%%%%%%
\subsection{Development of readout ASIC}
%%%%%%%%%%%%%%%%%%%%%%%%%%%%%%%%%%%%%%%%%%%%%

\begin{wrapfigure}{r}{0.4\columnwidth}
\centerline{\includegraphics[width=0.35\columnwidth]{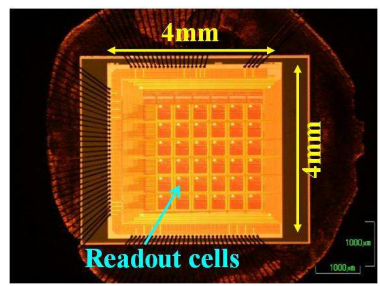}}
\caption{A picture of the readout ASIC.}
\label{fig:chip}
\end{wrapfigure}

We have developed the readout ASIC for the pair monitor. 
It is designed to count the number of hits to obtain the hit distribution 
on the detector. 
From our previous studies, 
the statistics for about 150 bunches is enough to extract the beam information 
on the detector. 
Therefore, the number of the hit is counted for 16 timing parts in one train, 
which corresponds to 167 (= 2670/16) bunches in the current nominal ILC design. 
The hit counts for each timing parts are read within the inter-train time (200 ms). 
A silicon pixel sensor with the thickness of about 200 \um~is assumed 
as a detector candidate, whose signal level is about 15,000 electrons.  
The readout ASIC is designed to satisfy these requirements. 

The readout ASIC consists of the distributor of the operation signals, 
shift register to specify a readout cell, data transfer to the output line, 
and 36 readout cells. 
A readout cell consists of the amplifier, comparator, 8-bit counter, 
and 16 count registers as shown in Fig. \ref{fig:logic}. 
They are aligned to 6 x 6 for the X and Y directions.
In the previous readout ASIC, 
MIM (Metal Insulator Metal) capacitors were not prepared at threshold block
due to mistake in the layout mask. 
Therefore, the signal line between pre-amplifier and differential amplifier 
was snapped \cite{fcal-proc}.

After modification of layout design, 
the prototype ASIC was produced with 0.25 \um~TSMC process 
as shown in Fig. \ref{fig:chip}. 
Its layout was made by Digian Technology Inc. \cite{digian}, 
and the production was done by the MOSIS Service \cite{mosis}. 
The chip size is $4 \times 4$ \mm, 
and the readout cell size is $400 \times 400$ \umum. 
In the readout cells, bonding pads are prepared to attach the sensors 
by bump-bonding. 
For the response test of the readout ASIC, 
the chip is covered with a PGA144 package.

%%%%%%%%%%%%%%%%%%%%%%%%%%%%%%%%%%%%%%%%%%%%%%%
\section{Response test of readout ASIC}
%%%%%%%%%%%%%%%%%%%%%%%%%%%%%%%%%%%%%%%%%%%%%%%
\begin{wrapfigure}{r}{0.45\columnwidth}
\centerline{\includegraphics[width=0.4\columnwidth]{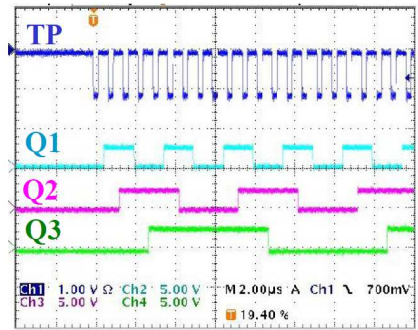}}
\caption{Output signals from the counter block.
TP shows the test-pulse, and Q1, Q2, and Q3 show the counter bit.
The hit count is output with Gray code.}
\label{fig:oscil}
\end{wrapfigure}

For the response test of the readout ASIC, 
the test system was constructed, based on the VME system. 
A GNV-250 module was used for the operation and data readout, 
which was developed as the KEK-VME 6U module. 
The readout ASIC is attached on the test-board, and connected to the GNV-250 module. 
Since a FPGA is equipped on the GNV-250 module, 
logic for data processing could be easily modified. 
To readout all the hit counts from 16 count registers in each pixel (36 pixels), 
we prepared a FIFO in the FPGA. 
All the hit counts are stored in it, then, they are sent to a computer.

At first, the response of the amplifier block was checked. 
The amplifier block consists of the pre-amplifier, threshold block, 
and differential amplifier as shown in Fig. \ref{fig:logic}.
The monitor output is prepared to check the internal signals 
after the pre-amplifier and differential amplifier.
In the previous ASIC, the signal from the pre-amplifier could not be sent to
the differential amplifier due to the problem in the MIM capacitor.
We could observe all the monitor outputs for new readout chip,
therefore, the amplifier block was confirmed to works correctly. 

\begin{wrapfigure}{r}{0.45\columnwidth}
\centerline{\includegraphics[width=0.4\columnwidth]{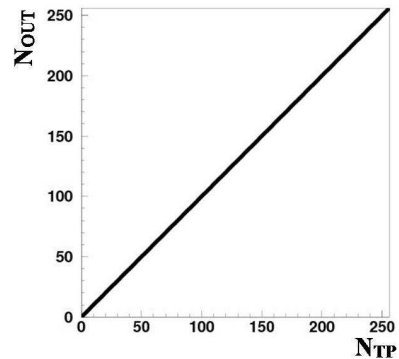}}
\caption{The relation between a number of the input pulse (N$_{\mathrm{TP}}$)
and that of the hit counts read from one of the count registers (N$_{\mathrm{OUT}}$),
which was obtained with about 1 MHz counter rate. 
No bit lost was observed in the data.}
\label{fig:data}
\end{wrapfigure}

For the next step, a function of the counter block was checked. 
Fig. \ref{fig:oscil} shows the output signals from the counter block, 
which was designed to use Gray code. 
Since the number of the hits was output correctly, 
the hit count was read from the count registers by a computer. 
Fig. \ref{fig:data} shows the relation between a number of the input pulse 
and that of the hit counts read from one of the count registers, 
which was obtained with about 1 MHz counter rate. 
It was confirmed that there is no bit lost in the data. 
%From these test results, 
%we can conclude that all the circuit block works correctly 
%without the snapping between the pre-amplifier and threshold block.

%%%%%%%%%%%%%%%%%%%%%%%%%%%%%%%%%%%%%%%%%%%%%
\subsection{Pair monitor with SOI technology}
%%%%%%%%%%%%%%%%%%%%%%%%%%%%%%%%%%%%%%%%%%%%%
For the next step, we plan to develop the pair monitor with SOI 
(Silicon On Insulator) technology. 
The SOI technology is the technique to electrically separate the transistors from Si layer. 
It realizes to prepare the sensor and readout ASIC on the same wafer without bump-bonding. 
Since we already developed the readout ASIC with usual CMOS technology, 
its design can be used for the readout circuit. 
In addition, the circuit is completely free from latch-up
because the device substrates are electrically separated each other.
We, however, still need a delicate study on the total ionization dose effect.
Even with a deep submicron process on a very thin Si layer, 
the devices are easy to be affected by a positive charge trapped in a BOX layer. 

This project was already started as collaboration with KEK and Tohoku university.
For the next production, only the readout ASIC will be developed 
without a sensor block to check its response independent of the sensor. 
The design was already fixed, 
and the first prototype will be delivered in 2009.

%%%%%%%%%%%%%%%%%%%%%%%%%%%%%%%%%%%%%%%%%%%%%
\subsection{Conclusions}
%%%%%%%%%%%%%%%%%%%%%%%%%%%%%%%%%%%%%%%%%%%%%
We developed the pair monitor for the beam profile monitor at ILC. 
The new readout ASIC was developed in 2008
which was modified to implement the MIM capacitor at the threshold block.
All the components were confirmed to work correctly by the response test. 
For the next step, we plan to develop the pair monitor with SOI technology. 
The first prototype will be developed in 2009.

\section{Acknowledgments}
The authors would like to thank all the member of the
FCAL collaboration \cite{fcal} for all their help.
This study is supported in part by the Creative Scientific Research Grant
No. 18GS0202 of the Japan Society for Promotion of Science 
and promotion of collaborative research programs in universities with KEK.

% ****************************************************************************
% BIBLIOGRAPHY AREA
% ****************************************************************************

\begin{footnotesize}
% IF YOU DO NOT USE BIBTEX, USE THE FOLLOWING SAMPLE SCHEME FOR THE REFERENCES
% ----------------------------------------------------------------------------

% ----------------------------------------------------------------------------

% IF YOU USE BIBTEX,
% - DELETE THE TEXT BETWEEN THE TWO ABOVE DASHED LINES
% - UNCOMMENT THE NEXT TWO LINES AND REPLACE 'Name_Of_Your_BibFile'

%\bibliographystyle{unsrt}
%\bibliography{Name_Of_Your_BibFile}
% example of Name_Of_Your_BibFile.bib
% @Article{Turcato:2006ch,
%      author    = "Turcato, M.",
%  collaboration = "ZEUS and H1",
%      title     = "Lepton flavour violation and charmonium physics at HERA",
%      journal   = "Nucl. Phys. Proc. Suppl.",
%      volume    = "162",
%      year      = "2006", 
%      pages     = "283-287",
%      SLACcitation  = "%%CITATION = NUPHZ,162,283;%%"
% }
% 
% @Unpublished{Gogitidze:2007du,
%      author    = "Gogitidze, N.",
%  collaboration = "H1", 
%      title     = "Prompt photons and particle momentum distributions at
%                   HERA", 
%      year      = "2007",
%      note    = "hep-ex/0701033",
%      SLACcitation  = "%%CITATION = HEP-EX 0701033;%%"
% }

\end{footnotesize}

% ****************************************************************************
% END OF BIBLIOGRAPHY AREA
% ****************************************************************************

\end{document}